\documentstyle[preprint,prd,aps]{revtex}

\begin{document}
\draft
%%%%%End of Preamble
%%%%Start of Text%%%%%%%%%%%%%%%%%%%%%%%%%%%%%%%%%%%%%%%%%%%%%%%%%%%%%%%
\preprint{
\vbox{
\halign{&##\hfil\cr
	& AS-ITP-97-11 \cr
	& hep-ph/9704364 \cr
	& March. 1997 \cr}}
}
\title{The $J/\psi$ Production Associated with a Hard Photon
and the Color-Octet Singnature in $e^+e^-$ Annihilation}

\author{Chao-Hsi CHANG$^{\S,\dagger}$, Cong-Feng QIAO$^{\S}$
and Jian-Xiong WANG$^{\S,\ddagger}$}
\address{ $^{\S}$ CCAST (World Laboratory), P.O. Box 8730,
 Beijing 100080, China.\footnote{Not mailing address for
C.-H. CHANG and J.-X. WANG.}}
\address{ $^\dagger$ Institute of Theoretical Physics,
 Academia Sinica, Beijing 100080, China.}
\address{ $^\ddagger$ Institute of High Energy Physics,
 Academia Sinica, Beijing 100039, China.}

\maketitle

\vspace{-5mm}
\begin{abstract}

$J/\psi$ production associated with a hard photon
in $e^+e^-$ annihilation, being of QED nature,
is invastigated thoroughly in the paper.
To show its influence on the observation of the color-octet
singnature in the $e^+e^-$ annihilation via $J/\psi$ inclusive
production, the cross sections of the $J/\psi$ productions through
different mechanisms at various energies are
compared quantatively by presenting them in figures
together. The contribution from the production associated
with a hard photon to the inclusive production of $J/\psi$ is
pointed out to be significant at the concerned energies, thus
the influence from it on the observation of
the color-octet signature should be dealt with carefully.

\vspace{3mm}
\noindent
{\bf PACS} numbers:  12.20.Ds, 13.65.+i, 12.20.Fv, 12.38.Qk

\end{abstract}

\vfill \eject

%\narrowtext

Since $J/\psi$ was discovered more than twenty years ago,
its hadronic production has been interesting people,
because of the problem being open.
Recently several progresses on the production are achieved.
One is about the calculations
of the fragmentation functions and the corresponding production mechanisms
for the double heavy meson (heavy quarkonia and $B_c$ meson
etc.) productions, i.e. in the framework of
perturbative QCD, the fragmentation
functions and the productions via the so-called fragmentation mechanisms
are realized to be calculable and the calculations are believed to
be reliable\cite{frag} . The second is about
tests of the predictions of the calculations
with experimental data\cite{data}.
As a result, a precise conflict between the
theoretial predictions and the Tevatron experimental data, i.e.
`$\psi' (\psi)$ surplus' puzzle, is confirmed\cite{plus}.
The third is that an interesting
solusion for the puzzle, the so-called `color-octet mechanism',
is proposed\cite{octet}. Due to the progresses, now the
interests in the problem and relevant ones
are freshened and spreading widely. To confirm the color-octet
mechanisms\footnote{In fact, in various color-octet
productions of $J/\psi$, the acting subprocesses are
different from each other, not only in the order
of the QCD in the strong coupling $\alpha_s$ but also in the behavior of the
propagators in the Feynman diagrams resposible for the subprocesses,
thus we think it is better to call a production with a different acting
subprocess as a different mechanism.} further
and to test the proposal, many suggestions appear.
For instance, investigations on all kinds of the mechanisms
in the $J/\psi$ inclusive production, including the color singlet ones
and the octet ones, at various energies and in various
processes, not only in hadronic colliders but also in $e^+e^-$ or
$ep$ colliders etc, emerge in preprints, letters
and achieved papers\cite{flem,ee,photon,cho1,octets,fixed,braaten,cho}.
%and achieved papers\cite{flem} -
%\cite{ee}\cite{photon}\cite{cho1}\cite{octets}\cite{fixed}\cite{braaten}
%\cite{cho}.
Moreover of them, being clean, the productions of $J/\psi$ through
$e^+e^-$ annihilation are specially emphasized by several authors
to confirm the octet mechanisms\cite{braaten,cho,ktchao},
%to confirm the octet mechanisms\cite{braaten} =
%\cite{cho}
%\cite{ktchao}.
Whereas, the $J/\psi$ production associated with a hard photon via $e^+e^-$
annihilation, deplicted by the Feynman diagram Fig.1.a,
is less considered in literature. For convenience, throughout the paper
we will call it as a hard-photon production (or mechanism) latter on.
In fact, the hard-photon production should be catalogued in color
singlet mechanism and is of QED nature essentially.
Based on a rough order-estimate of the couplings and propagators
to the corresponding Feynman diagrams, one can be sure
that it contributes to the inclusive $J/\psi$ production
in energetic $e^+e^-$ annihilation in a certain amount,
although it is one order higher in the coupling $\alpha$
than `others'.
Therefore precisely to calculate the process versus the others,
including the comparatively well-studied mechanisms such as
the color-octet ones (Fig.1.b) and the color-singlet ones (Fig.1.c),
at various energies becomes interesting.
In the paper we are to do the calculations
to see how significant is the contribution from each mechanism.
Finally a conclusion is reached
that in $e^+e^-$ annihilation the contribution from the hard-photon
mechanism to the inclusive production of $J/\psi$ is
very significant at the concerned energies, and its affects on
detecting the
singnature of the color-octet mechanism are great.
%%%%strongly.
In the $e^+e^-$ annililation
all the concerned color-octet mechanisms
and the color-singlet mechanisms, except the
hard-photon one, are of `s-channel annihilation',
so the propagator of the virtual photon from the $e^+e^-$ annililation
plays a role of a suppression factor, especially,
when the annihilation happens at a high energy.
Thus the interesting color-octet signature
is expected to be observed better at comparatiely low energies
such as at TRISTAN, CESR and BEPC etc, rather than
at high ones such as at LEP and SLC.
Furthermore, at a relatively high energy, e.g. $\sqrt s \geq m_Z$,
in the production more acting mechanisms are involved,
so the production is more complicated, that
we will discuss them elsewhere\cite{cchw}.

Theoretically, except the binding factor of $c\bar c$ in
the $J/\psi$, the rest parts of the $J/\psi$ production
associated with a hard photon in $e^+e^-$ annihilation
are of QED. The process may be computed
accuratly with QED Feynman diagrams.
To the lowest order, as depicted in Fig.1.a
it is quite a general feature that
the hard photon in the process couples to an
electron line, and so does the $J/\psi$ but
through a virtual photon indirectly.
Here, merely for simplisity, in Fig.1.a only one of the Feynman diagrams for
the process is presented, but when we calculate the process,
complete set of the diagrams (here only one else, with the two
photon lines crossed, should be added) responsible for the process
are considered.
Throughout the paper and for all processes
the same simplification in drawing
Feynman diagrams is taken.
Namely Feynman diagrams for a concerned process
are presented in figures typically.
In fact, in Fig.1.a. the virtual photon is the same
as that of the $J/\psi$ production in $e^+e^-$
annihilation at resonance and that in the decay $J/\psi \to
e^+e^-$, i.e.the momentum of the virtual photon is just
`on-shell' of $J/\psi$, thus it is not a suppression factor.
Note here that the so-called electromagnetic fragmantation
approach (EMFA)\cite{flem} works well only at much high energies
($\gg m_{J/\psi}$) and, as mentioned above,
we constrain ourselves to consider all kinds of production at
the energies of BEPC, CESR and TRISTAN\footnote{The energies, at which
experimental data were taken at BEPC, CESR and TRISTAN,
are chosen to compute the production.
Moreover, in general, there is always an additional
$Z$ boson exchange diagram,
which is obtained by replacing each virtual
photon once for the diagrams Fig.1.a-1.c with a virtual $Z$ boson.
Whereas at the comparatively low energies,
being much lower than $m_Z$, the contributions from the $Z$ boson
exchange diagrams are tiny (small) so we ignore them throughout
the calculations.}, which even may compare
with $m_{J/\psi}$, so EMFA
is not applicable here. Furthermore there would be no
advantages in calculations of the process if we had adopted
EMFA no matter how it had worked. Therefore here we do not adopt it.
Experimentally, if the associated hard photon in the process can be
identified well, the process will be measurable
exclusively with certain accuracy, hence
the theoretical calculations will become testable by experiments
directly.

In fact, there is additional color-singlet mechanism
$e^+e^-\to J/\psi +f+\bar f$ in the $e^+e^-$ annihilation,
where $f, \bar f$ denotes
a pair of fermions in `flavor' $f$
(quarks or leptons), thus this mechanism involves many chennals
with various fermion pairs $f\bar f$. For each channel
with a specific pair $f, \bar f$,
the corresponding Feynman diagrams
are obtained in such a way, that
via a virtual photon line indirectly,
the $J/\psi$ couples to a line of the final
fermion $f$ or $\bar f$ or of the initial electron or positron
in turn to its `skeleton' diagram, which just is
the annihilation processe $e^+e^- \to f\bar f$.
Although they are of QED essentially too and also less considered
in literature, because they are one order higher in $\alpha$
than the considered hard-photon one,
we will study them elsewhere\cite{cchw}.

The corresponding amplitude for the hard-photon
production may be written down immediately, and
with a straightforward calculation, the
differential cross section is obtained:
\begin{equation}
\displaystyle \frac{d\sigma}{dt}=\frac{32\pi\alpha^3|R_S(0)|^2}
{3M^3S^2} [\frac{2M^2 s}{tu}+\frac{t}{u}+\frac{u}{t} ],\\[2mm]
\end{equation}
where
$$ s=(p_1+p_2)^2; \;\; t=(k-p_1)^2; \;\; u=(P-p_1)^2.$$
For simplicity, here in the formulae the electron mass is
ignored, whereas in the numerical calculations we keep it so as to
avoid the `colinear divergence'
in the beam direction for the differential cross section.
As the wave function at the origin $|R_S(0)|$ appearing here is
exactly as that appearing in
the width of the decay $J/\psi \to e^+e^-$, i.e.\
\begin{equation}
\displaystyle \Gamma(J/\psi \to e^+e^-) = \frac{16\alpha^2}{9M^2}
|R_S(0)|^2,\\[2mm]
\end{equation}
in numerical calculations, we with eq.(2)
use the experimental width for the decay $J/\psi \to e^+e^-$
as an `input' instead of the wave function. Namely
\begin{equation}
\displaystyle \frac{d\sigma}{dt}=
\frac{6\pi\alpha\Gamma^{exp.}(J/\psi \to e^+e^-)}
{MS^2} [\frac{2M^2 s}{tu}+\frac{t}{u}+\frac{u}{t} ].\\[2mm]
\end{equation}
In the above way to determine the requested wave
function at origin, the strong interaction corrections
on the wave function have been included.
With eq.(3), the numerical values for
the differential cross secton as a function of
$t$, or with a straightforward calculation
the differential cross secton $d\sigma/d\cos\theta$ as a function of
the angular $\cos\theta$ ($\theta$ - the angular between the incoming
beam direction and that of the outgoing $J/\psi$) at a given CMS energy,
and the total cross section as a function of the CMS energy
by integration of eq.(3) may be calculated precisely.
Here we choose the CMS energies
$\sqrt S= 4.03 GeV; 10.6 GeV; 64.0 GeV$, where quite a lot of data
were taken,
for differential cross sections to do the numerical
calculations and plot the results in Fig.2.a-2.c, and
the total cross section in Fig.3, respectively.
In order to compare the hard-photon mechanism with the others
in the contribution to the inclusive production of $J/\psi$
in different aspects, and to show its characteristics concisely,
here only those important ones: color octet ones depicted in
the Feynman diagram Fig.1.b and a color siglet one in the Feynman
diagram Fig.1.c are selected as the `typical others'.
While for the mechanisms of Fig.1.b and
Fig.1.c, the request formulae in the calculations are refered from
literatures \cite{cho1} and \cite{cho}. Namely when we calculate the
contributions, the formulae for the color-octet are quoted
from \cite{cho1}, and the ones for the color-siglet from \cite{cho}.
For convenience to compare the different mechanisms we plot the results
of the hard-photon one (Fig.1.a) and the considered `typical others'
(Fig.1.b, Fig.1.c) into the same figure.

It is easy to realize that the production of the hard-photon
is very different from the others because it is a `t,u-channel'
process (see Fig. 1.a.) whereas the others, as mentioned above, are `s-channel'
ones (see Fig. 1.b-1.c). In general, the differential cross section of
the production, as long as it is a `s-channel' process,
can certainly be formulated as the follows:
\begin{equation}
\displaystyle \frac{d\sigma}{dEd\cos\theta}=
S(E)[1+A(E)\cos^2\theta] \\[2mm]
\end{equation}
which is emphasized by the authors of \cite{braaten} and
where $E$ and $\theta$ are the energy of the produced $J/\psi$
and the angle between the directions of the
$J/\psi$ and beam at CMS respectively,
whereas, for the hard-photon production, which is of a `t,u-channel,
the differential cross section cannot be formulated as eq.(4).
Therefore it is not a good way to see the influences
from the later (`t,u-channel) on the fermer
(`s-channel') with the formulation eq.(4). If insisting on
the formulation eq.(4) to compare them
with each other, we should rewrite the differential cross section
eq.(1) into a one,
depending on $\cos\theta$ explicitely:
\begin{equation}
\displaystyle \frac{d\sigma}{dEd\cos\theta}=\delta(E-E_{max})
\frac{32\pi\alpha^3|R_S(0)|^2}{3M^3S(1-r)\sin^2\theta}
[(1+r)^2+(1-r)^2\cos^2\theta], \\[2mm]
\end{equation}
where $r=M^2/S$. If the factor $\sin^2\theta$ in the denominator
of the above differential cross section were ignored,
the equation eq.(5) would turn
to the formulation of eq.(4), and it would be
interesting to note that the coefficient before $\cos^2\theta$ of the second
term in the squared braket of eq.(5) would take a positive
sign, as that of color-octet ones \cite{braaten}.

To see the influences of the hard-photon production onto
the signature of the color-octet mechanisms in the formulation eq.(4),
we try to define a `equavelent' factor
$\hat A_{eqv}(E,\theta)$
by differential cross sections:
$$\displaystyle \hat A_{eqv}(E,\theta)
\;=\;\frac{1}{\cos^2\theta}\cdot[\frac
{\frac{d\sigma}{dEd\cos\theta}}
{\frac{d\sigma}{dEd\cos\theta}|_{\cos\theta=\frac {\pi}{2}}}-1]. $$
For `s-channel' processes, such as the color-octet ones, the factor
$\hat A_{eqv}(E,\theta)$ does not depend on the $\theta$ at all,
and turns to the factor $A(E)$ in eq.(4) exactly, but for the
hard-photon production, being of a `t,u-channel' one, does depend on $\theta$
but in the sense of the formulation eq.(4) we may consider it as
an equavalent factor of $A(E)$. Let us now
take $\sqrt S=10.6 GeV$ as an example, to show the behavior
of the $\hat A_{eqv}(E,\theta)$ in Table I.
\begin{center}
{\bf Table I. The Factor $\hat A_{eqv}(E,\theta)$ at $\sqrt S=10.6 GeV$.}

\vspace{3mm}
\begin{tabular}{|c|c|c|c|c|c|c|c|c|c|}\hline
$\theta$ & $\pi/2$ & $\pi/3$ & $\pi/4$ & $\pi/5$
& $\pi/6$ & $\pi/7$ & $\pi/8$ & $\pi/9$ & $\pi/10$ \\[2mm]
\hline
$\hat A_{eqv}(E,\theta)$ & 1.710 & 2.28 & 3.42 & 4.95
& 6.84 & 9.08 & 11.68 & 14.62 & 17.91 \\[2mm]
\hline
\end{tabular}
\end{center}
For reference, from \cite{braaten}
we have $\hat A_{eqv}(E,\theta)\equiv A(E)=
\simeq 0.6\sim 1.0$ for color-octet
productions (Fig.1.b) and $\hat A_{eqv}(E,\theta)\equiv
A(E)=-0.84$ for the color-singlet
one (Fig.1.c), whereas here from Table I.
$\hat A_{eqv}(E,\theta)\geq 1.7$.

As a matter of fact,
the best way to show the characters of the hard-photon production of
$J/\psi$ and to compare it with
the `other' productions quantatively is to present their total
and differential cross sections in figures respectively.
Therefore we are
doing so now.

In Fig.2.a-2.b the differential cross sections, $d\sigma/d\cos \theta$
verus $\cos \theta$ for the productions depicted
by Fig.1.a-1.c at the CMS energies
$\sqrt S= 4.03 GeV; 10.6 GeV; 64.0 GeV$ are plotted
respectively. One can from the figures see
that to the inclusive production of $J/\psi$ via $e^+e^-$ annihilation,
the hard-photon production contributes a dominant fraction
over the others' when the $J/\psi$ goes out near the beam direction,
and still a significant fraction when the $J/\psi$ goes out
in the direction near the direction perpendicular to the beam direction.
Namely even in the direction of the $J/\psi$
perpendicular to the beam direction,
the hard-photon one still contributes a fraction
not smaller than the greatest one amoung the considered `others'
by a factor $3$, although the greatest one among the
considered ones is alternated with the CMS energy is increasing.
We should note here that {\bf at the considered
CESR energy $\sqrt s=10.6 GeV$, the hard-photon mechanism} (Fig.1.a)
{\bf has an
angular distribution in a shape similar to that of the color-octet
mechanisms} (Fig.1.b), {\bf but different
from that of the color-singlet one} (Fig.1.c)
{\bf that makes its $\hat A_{eqv}(E,\theta)$ positive as that of color-octet
ones, whereas that of the color-singlet one is negative.} (see Table.I)
To show the energy depedence precisely,
we plot the total cross sections versus the CMS
energies for the various mechanisms in Fig.3. Considering the existing
experiment detector(s), it is possible to measure the hard photon
and/or the produced $J/\psi$ if the photon as well as the $J/\psi$,
goes out not very close to the beam direction or if
the production happens at a comparatively low energy
that the $J/\psi$ moves not very fast no matter the direction
how close to the beam. Thus in Fig.3
we plot two curves for the hard-photon mechanism: one (thick
solid line) is with a cut on the outgoing anglar of the $J/\psi$ and the
other (thin solid line) without any cut respectively.
It is easy to see that, when no cut is made,
the hard-photon mechanism contributes to the inclusive
production of $J/\psi$ dominantly over all the others,
and when a cut $20^0\leq \theta \leq 160^0$ in angulars
is put on, the hard-photon mechanism does not dominate over all
the others but in the considered
CMS energy region it is comparitable to that of the biggest one
among the others considered here.
>From the figure (Fig.3), one may also see some interesting
aspects for the `other' mechanisms. For instance, the color-singlet
one (Fig.1.c) is smaller than those of the color-octet ones
(Fig.1.b) in the energy region $\sqrt S \leq 12 GeV$ but becomes
greater in the energy region $\sqrt S \geq 12 GeV$. All the results
shown here are understandable qualitatively: As for the hard-photon one,
being a `t,u-channel' production, there is an enhencement factor
especially, when the t-channel or the u-channel virtual electron line
appoaches to the electron mass-shell
(it is the reason its differential cross section
becomes very large in the beam direction), to
compare with s-channel prosses when $\sqrt S\gg
M^2$. Besides an overall suppression factor due to one order
higher in $\alpha$ than the `others', the hard-photon production
gains a factor $\alpha_s^{-2}$ to compare with the color-singlet
one (Fig.1.c) and it gains a factor $\alpha_s^{-1}v^{-4}$,
where $v$, being small,
is the typical relative velocity of the quarks in a heavy quarkonium
$J/\psi$ to compare with the color-octet one (Fig.1.b).
In summary of the factors, it is the reason why
the cross sections of the concerned mechanisms
present so complicated feature in Figs.2.a-c and Fig.3.

Before drawing a conclusion, we should note two points:
i). In the plots the values of the color-octet matrix elements
for the color-octet mechanisms are taken from the
determination by fitting the Tevatron data\cite{cho1},
and it seems that they are greater than those determined from
photoproduction and fixed target experiments\cite{photon},\cite{fixed},
therefore here it may be overestimed a little for the color-octet ones.
ii). In hadronic and photonic productions of $J/\psi$,
there is a similar mechanism to the hard-photon one emphasized
in the paper, we will discuss its effects elsewhere\cite{cchw}.

In conclusion, the hard-photon production of $J/\psi$
itself is an interesting physics in energitic
$e^+e^-$ annihilation because of its less studying.
Its feature is quite different from others if using
the factor
$\hat A_{eqv}(E,\theta)$
to observe (Table I).
If one would like to observe the signature
of the color-octet mechanism in $e^+e^-$ annihilation
merely in the way as suggested by \cite{braaten},
it may not be successful, because
the experimental error may not be deducted very well and
the gap of $\hat A_{eqv}(E,\theta)$ for the color-octet one
and the hard-photon one is not so great,
for instance, at $\sqrt S=10.6 GeV$ it
has only 0.7, i.e. from 1.0 to 1.7.
At the energies, such as those of BEPC,
CESR and TRIESTON, if one may separate the contribution of the
hard-photon mechanism precisely in the experimental observation
in the $e^+e^-$ annihilation at certain level,
to observe the signature of the color-octet mechanisms and with the
suggested way by \cite{braaten}
may be practable. In principle it is accessible to separate
the contribution of the hard-photon mechanism precisely
by means of a precise exclusive
measurement of the production, i.e., not only to measure the produced
$J/\psi$ but also the associated hard photon
(with experimental tagging techniques). Therefore,
in order to study the color-octet mechanisms
and to observe the color-octet signature in $e^+e^-$ annihilation
without obscurity, the experimental data of the inclusive production
of $J/\psi$ should be `cleaned up' in certain level, i.e.
the contribution of the hard-photon production should be
separated some in advance.

\vspace{2cm}
{\large \bf Acknowledgement} This work was supported in part by
the National Science Foundation of China and the Grant LWTZ-1298 of Chinese
Academy of Science.

\vfill\eject

%%%%%%%%%%%%%%%%%%%%%%%%%%%%  FIGURE CAPTIONS  %%%%%%%%%%%%%%%%%%%%%%%%%%%%%%

\figure{Fig.1.a A typical Feymann diagram for the production of $J/\psi$
with a hard photon via $e^+e^-$ annihilation.}
\figure{Fig.1.b A typical Feymann diagram of some color-octet mechanisms
for the production of $J/\psi$ via $e^+e^-$ annihilation.}
\figure{Fig.1.c A typical
diagram of a typical color-singlet mechanism
for the production of $J/\psi$ via $e^+e^-$ annihilation.}

\figure{Fig.2.a The diffrential cross sections $d\sigma/d\cos \theta$
of the production $J/\psi$ versus $\cos \theta$ for the various
mechanisms at the CMS energy $\sqrt s=4.03 GeV$ (BEPC).}
\figure{Fig.2.b The same as Fig.2.a but at the CMS
energy $\sqrt s=10.6 GeV$ (CESR).}
\figure{Fig.2.c The same as Fig.2.a but at the CMS
energy $\sqrt s=60.0 GeV$ (TRISTAN).}

\figure{Fig.3. The total cross sections
of the production $J/\psi$ versus the CMS energy $\sqrt S$ for the
various mechanisms: the thin solid curve presents that for the mechanism
with a hard photon without any cut; the thick solid one presents that
for the same mechanism but with a cut $20^0\leq \theta \leq 160^0$.}

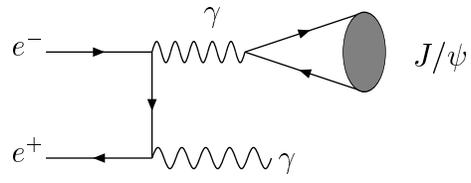
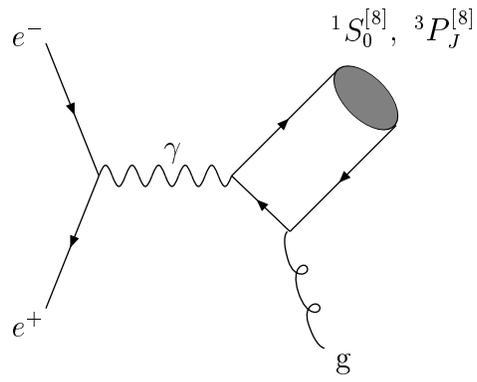
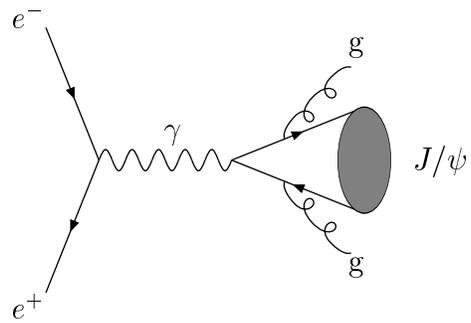
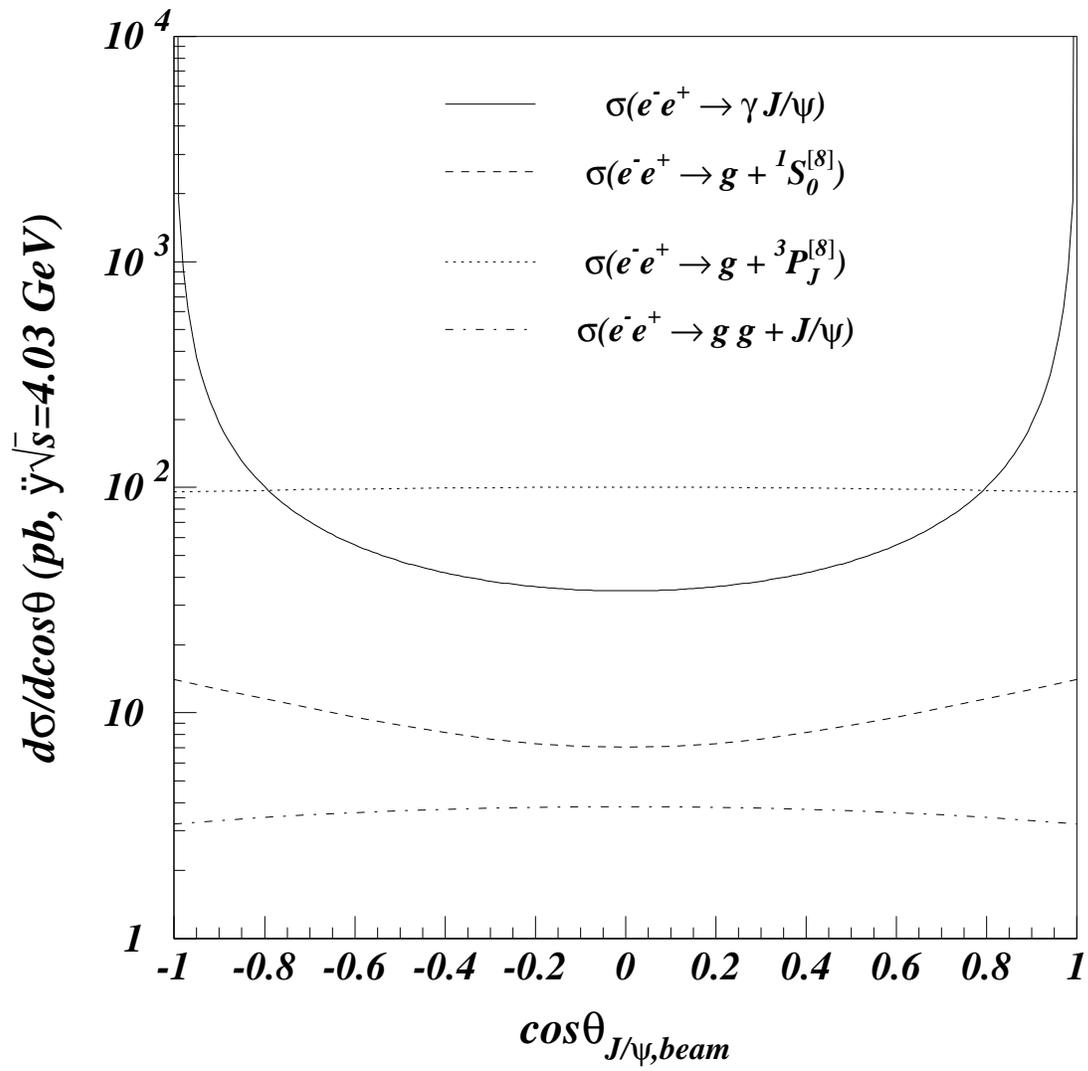
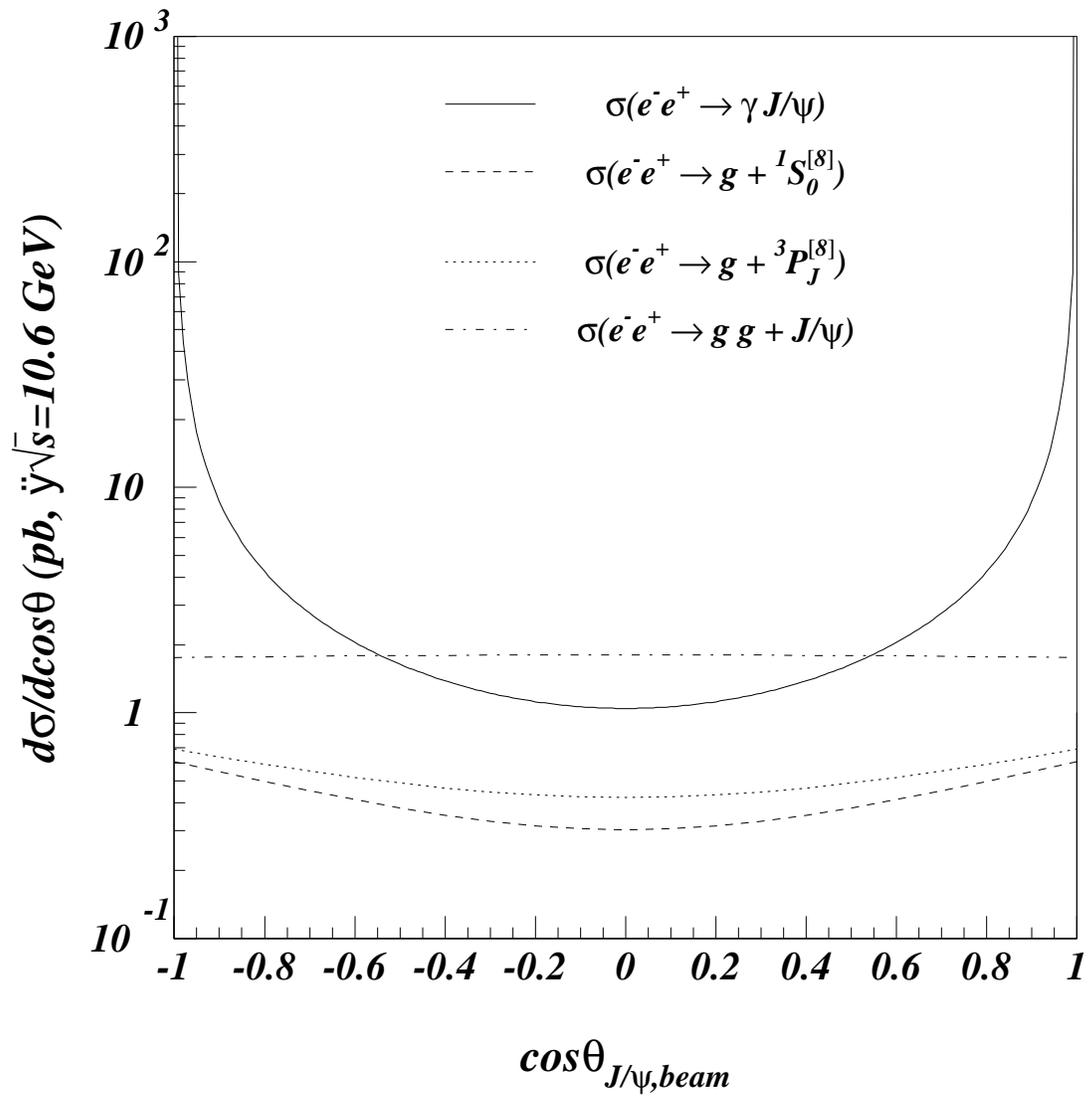
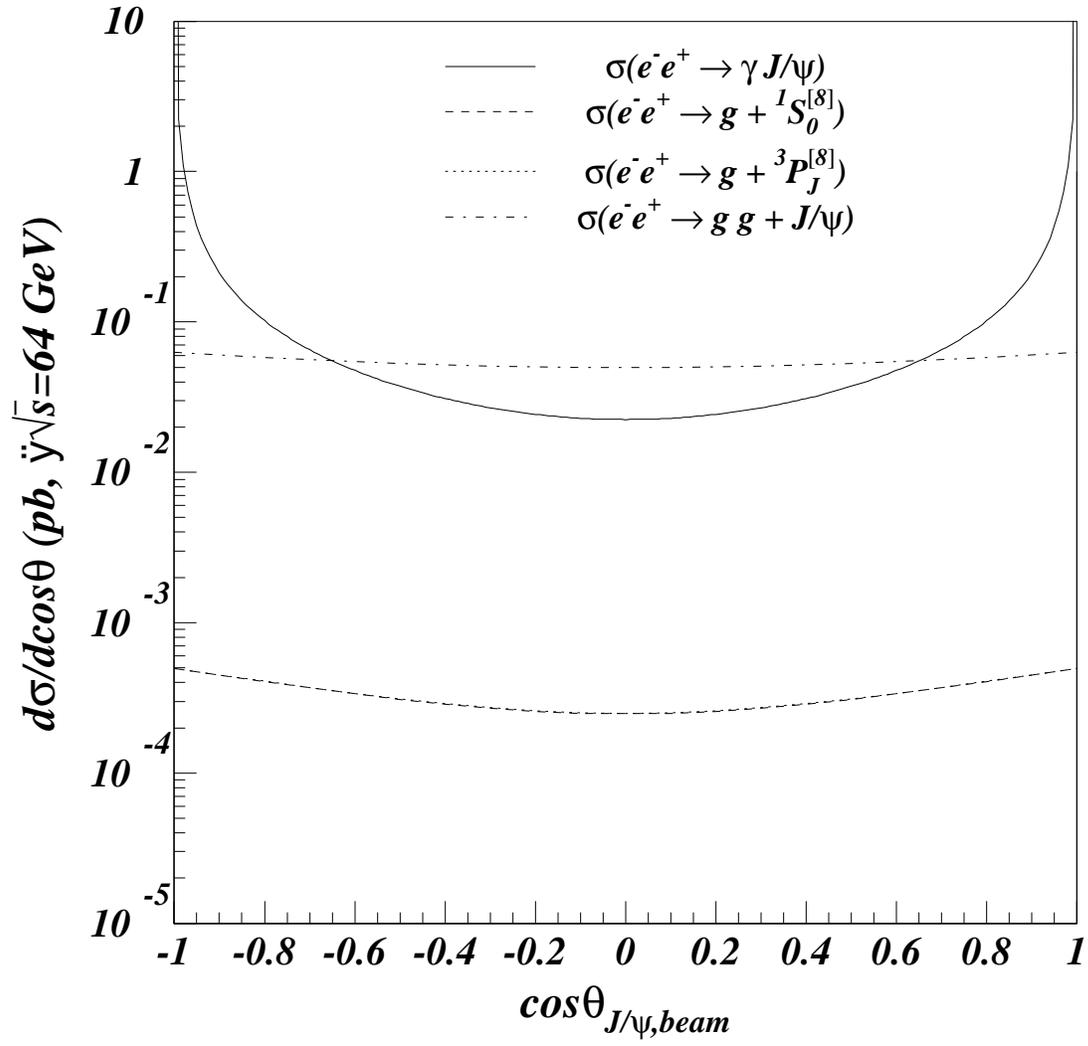
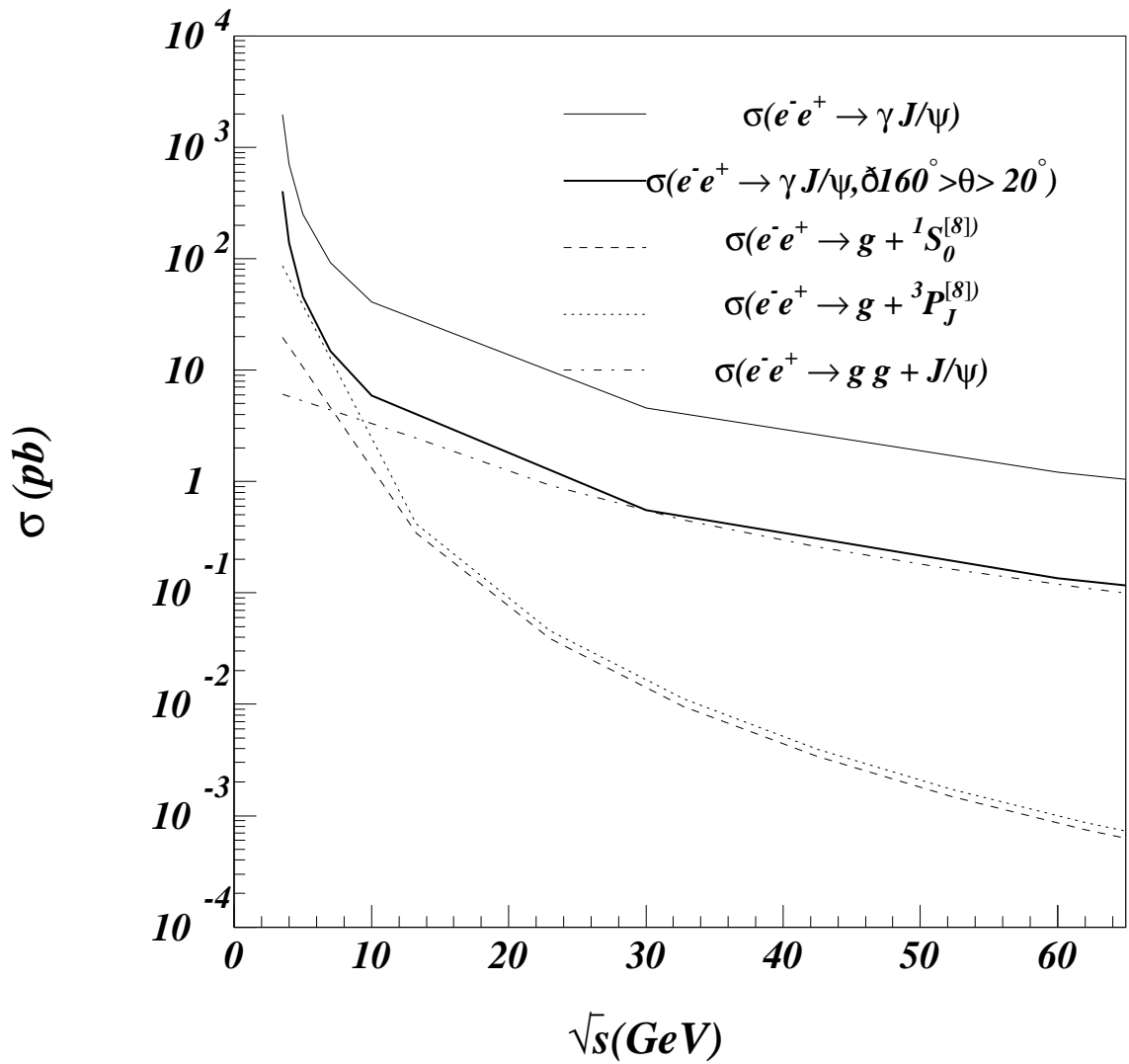
\end{document}